\documentstyle[12pt]{article}
\begin{document}
%\draft
%\title{EFT and NN Scattering}
\begin{center}
{\bf Chiral Perturbation Theory Approach to NN Scattering Problem}
\end{center}
%\medskip
\begin{center}
{\bf J. Gegelia\footnote{e-mail address:
gegelia@daria.ph.flinders.edu.au}} \\
\smallskip
{\it  School of Physical Sciences, Flinders University of South 
Australia,}

{\it Bedford Park, S.A. 5042, Australia}  
\end{center}
 
\medskip
%\end{center}
%\author{J.Gegelia${ }^a$\footnote{e-mail address:
%gegelia@daria.ph.flinders.edu.au}}
%\address{School of Physical Sciences, Flinders University of South 
%Australia, \\ Bedford Park, S.A. 5042, Australia}
%\end{center}
%\begin{center}
%{\it School of Physical Sciences, Flinders University of South Australia,}
%\end{center}
%\medskip
%\medskip
%\medskip
\date{\today}
%\maketitle

\begin{abstract}
   It is shown that chiral perturbation theory (in its original form by
Weinberg) can describe 
$NN$ scattering with
positive as well as negative effective range. Some issues connected with
unnaturally large $NN$ ${ }^1S_0$ scattering length are discussed.
\end{abstract}

\medskip
\medskip
\medskip

\medskip
\medskip
%\newpage
The chiral perturbation theory approach to the low-energy purely pionic
processes \cite{wein1979} has been generalised for processes involving
an arbitrary number of nucleons 
\cite{weinberg1}, \cite{weinberg2}. Weinberg pointed out that for the
$n$-nucleon problem the power counting
should be used for the ``effective potentials''  and not for the full
amplitudes. The effective potential is defined as a sum of time-ordered
perturbation theory
diagrams for the $T$-matrix excluding those with purely nucleonic intermediate
states.  

 The full $S$-matrix can be obtained by solving a Lippmann-Schwinger equation
(or Schr\" oedinger equation) with this effective potential in place of the
interaction Hamiltonian, and with {\it only} $n$-nucleon intermediate states 
 \cite{weinberg1}.

  Using renormalization points also characterised by momenta of order of
external momenta $p$ or less, the ultraviolet divergences that arise in
calculations using  effective
theory are absorbed into the parameters of the Lagrangian.
After renormalization, the effective cut-off is of order $p$  \cite{weinberg2}.
The
Lagrangian is rich enough to contain all possible terms which are allowed by
assumed symmetries, so all necessary counter-terms are present in the 
Lagrangian.

There has been much recent interest in the chiral perturbation theory approach
to nucleon-nucleon scattering problems. While some papers concern  different
constructive calculational and conceptual problems  
the authors of \cite{cohen1,phillips1,scaldeferri,unp,beane} 
came to the conclusion that Weinberg's approach to EFT has severe
problems. These arise in the
description of interactions with positive
effective range. Moreover they conclude ``there is no effective field theory of
$NN$ scattering with nucleons alone'' and the inclusion of pionic degrees of
freedom does not solve the problem \cite{beane}.
Below it is shown that the Maryland group encountered problems because some
features of EFT approach were not taken into account consistently. Performing
correct
renormalization one sees that  the above discouraging conclusions are
misleading and nothing is wrong with the EFT (chiral perturbation theory)
approach to the nucleon-nucleon scattering problem.

After realising that EFT does not suffer from fundamental problems one is still
left with the challenging problem \cite{kaplan} that unnaturally large
scattering
length of ${ }^1S_0$ wave nucleon-nucleon scattering restricts the validity
of EFT to very small values of energy (external
momenta). The use of the correct renormalization procedure leads one to the
natural solution of this problem via exploiting the freedom of the choice of
normalization condition. 

\medskip
\medskip
\medskip
%\section{Explicit calculations}

The effective non-relativistic Lagrangian for  very low
energy EFT, when the pions are integrated out is given by \cite{kaplan} 
$$
{\cal L}=N^{\dagger}i\partial_tN-N^{\dagger}{\Delta \over 2M}N-
{1\over 2}C_S\left( N^{\dagger}N\right)^2-{1\over 2}C_T
\left( N^{\dagger}\mbox{\boldmath $\sigma$}N\right)^2
$$
\begin{equation}
-{1\over 2}C_2\left( N^{\dagger}\Delta N\right)
\left( N^{\dagger}N\right)+h.c.+...
\label{e1}
\end{equation}
where the nucleonic field $N$ is a two-spinor in spin space and a two-spinor in
isotopic spin space and $\mbox{\boldmath $\sigma$} $ are the Pauli matrices
acting on spin indices. $M$ is
the mass of nucleon and the ellipses refer to additional 4-nucleon operators
involving two or more derivatives, as well as relativistic corrections to the
propagator. $C_T$ and
$C_S$ are couplings introduced by Weinberg \cite{weinberg1,weinberg2}, they are
of dimension $(mass)^{-2}$ and $C_2$ is of the order $(mass)^{-4}$.  

The leading contribution to the 2-nucleon potential is

\begin{equation}
V_0\left( {\bf p},{\bf p'}\right)=C_S+
C_T\left( \mbox{\boldmath $\sigma$}_1,\mbox{\boldmath $\sigma$}_2\right).
\end{equation}
In the ${ }^1S_0$ wave it gives:
\begin{equation}
V_0\left( {p},{p'}\right)=C
\end{equation}
where $C=C_S-3C_T$.
If we define $C_2\equiv {C\over 2\Lambda^2}$ (where $\Lambda$ is a parameter of
dimension of mass) the next to leading order contribution to the 2-nucleon
potential in the ${ }^1S_0$ channel takes the form: 

\begin{equation}
V_2\left( {p},{p'}\right)=C_2\left({p}^2+{p'}^2\right)=
C\left( {{p}^2+{p'}^2\over 2\Lambda^2}\right)
\end{equation}

Formally iterating the potential $V_0+V_2$ using the Lippmann-Schwinger equation
one 
gets for on-shell ($E=p^2/M$) $s$-wave 
$T$-matrix \cite{unp}: 
 
\begin{equation}
{1\over T(p)}={\left( C_2I_3-1\right)^2\over C+C_2^2I_5+
p^2C_2\left( 2-C_2I_3\right)}-I(p),
\label{2}
\end{equation}

\begin{equation}
I_n=-M\int {d^3k\over (2\pi )^3}k^{n-3}; \ \ 
I(p)=M\int {d^3k\over (2\pi )^3}{1\over p^2-k^2+i\eta}=I_1-{iMp\over 4\pi},
\label{3}
\end{equation}
where $p$ is the on-shell momentum and $I_1$, $I_3$ and $I_5$ are divergent
integrals.

The authors of the paper \cite{beane} carried the renormalization a certain
distance without specifying a regularization scheme, choosing as renormalised
parameters the experimental values of the scattering length $a$ and the
effective range $r_e$. $C$ and $C_2$ were fixed by demanding that  
\begin{equation}
{1\over T(p)}=-{M\over 4\pi}\left( -{1\over a}+
{1\over 2}r_ep^2+O\left( p^4\right)-ip\right)
\label{cut5}
\end{equation}
Expanding (\ref{2}) in powers of $p^2$ and comparing with (\ref{cut5}) we see
that imaginary parts agree and equating coefficients of terms of order $1$ and
$p^2$ we get:
\begin{equation}
{M\over 4\pi a}={\left( C_2I_3-1\right)^2\over C+C_2^2I_5}-I_1
\label{cut6}
\end{equation}
and 
\begin{equation}
{Mr_e\over 8\pi}=\left( {M\over 4\pi a}+I_1\right)^2\left[{1\over 
\left( C_2I_3-1\right)^2I_3}-{1\over I_3}\right]
\label{cut7}
\end{equation}
Rewritten in terms of $a$ and $r_e$ the scattering amplitude has the form:
\begin{equation}
Re\left( {1\over T(p)}\right)={M/(4\pi a)-p^2I_1A\over 1+p^2A}
\label{cut8}
\end{equation}
with
\begin{equation}
A\equiv {Mr_e\over 8\pi}\left( {M\over 4\pi a}+I_1\right)^{-1}
\label{cut9}
\end{equation}
If one uses sharp cut-off to regularize the divergent integrals then after
removal of cut-off one gets the finite result ($I_1\to\infty$, $A\to 0$ and
$I_1A\to {Mr_e\over 8\pi}$):
\begin{equation}
{1\over T(p)}=-{M\over 4\pi}\left[ -{1\over a}+
{1\over 2}r_ep^2-ip\right]
\label{cut10}
\end{equation}
But one can remove regularization only if $r_e\leq 0$. 
The problem is that one cannot solve $C_2$ from (\ref{cut7}) if $r_e$ is
positive and cut-off $l \to \infty$ (one gets a quadratic equation for $C_2$
which has no real solutions). The fact that one cannot take the
cut-off to infinity and still obtain positive
effective range is not surprising. Wigner's theorem states that if the potential
vanishes beyond range $R$ then for phase shifts we have \cite{wigner}:
\begin{equation}
{d\delta(p)\over dp}\geq -R+{1\over 2p}sin(2\delta(p)+2pR).
\label{cut11}
\end{equation}
From this one can derive \cite{phillips1}:
\begin{equation}
r_e\leq 2\left[ R-{R^2\over a}+{R^3\over 3a^2}\right]
\label{cut12}
\end{equation}
So, Wigner's theorem (derived from physical principles of causality and
unitarity) states that the zero range potential cannot describe positive
effective range. Consequently as far as the potential $V_0+V_2$ (with removed
cutoff)  is zero-range it
fails to describe actual nucleon-nucleon scattering with positive effective
range. 
Does it mean that EFT (with removed cut-off) fails to describe $NN$ scattering
with positive effective
range? The answer is {\it no}.

One should remember that although (\ref{2}) was obtained from the quantum
mechanics, it was written as an approximate expression of the QFT scattering
amplitude. Although in 
\cite{unp} it was shown that one can make (\ref{2}) finite
non-perturbatively, renormalizing only two parameters, one should remember that
our problem is an approximation to the effective field theory.  Thus to
renormalize (\ref{2}) 
consistently one should act in accordance with rules of EFT. In particular, one
should remove all divergences  by subtracting all divergent integrals or taking
into account contributions of counter-terms. 
If one takes the inverse of (\ref{2}) and expand in $C$ and $C_2$, one finds
that
this expansion contains increasing powers of $p^2$ with divergent coefficients.
To make the amplitude
finite, one has to include contributions of infinite number of counter-terms
with
number of derivatives growing up to infinity, or more simply subtract all
divergent integrals at some value of external momenta. One could think that this
fact makes the theory completely un-predictive, but it does not. Let us remind
that Weinberg's power counting applies to the  renormalised quantities, i.e.
after inclusion of contributions of counter-terms. The theory possesses the
predictive power because we expect that {\it renormalized} higher dimensional
couplings are heavily suppressed.  

The non-perturbative finiteness of the above potential model is not a
feature of the original field theory. In addition one hardly can expect that the
next approximations to the potential lead to non-perturbatively finite
results. So the above given renormalization scheme is not the one one should
follow.   Furthermore the power counting was performed in the renormalized
theory, so one can neglect
the contributions of higher order terms into physical quantities (like
scattering length and effective range) only in this theory. If one is working in
terms of regularized integrals and ``bare'' coupling constants, then one can not
neglect the contributions of higher order terms into effective range (or into
any other physical quantity) when the cutoff parameter is removed: there are
contributions of an infinite number of terms with more and more severe
divergences. So, the equations (\ref{cut6})-(\ref{cut7}) are not reliable and
hence being unable to solve $C_2$ in (\ref{cut7}) in the removed cut-off
limit does not mean that EFT is incapable of describing processes with a
positive effective range.  
 The only correct way of
renormalizing (\ref{2}) (consistent with QFT) is the approach by subtraction of 
integrals. Otherwise one should refer to the original effective field theory:
introduce all
counter-terms and sum all relevant {\it renormalised} diagrams up. Note that the
above non-perturbative expression for amplitude $T$ is nothing else than the sum
of infinite number of perturbative diagrams. It is easy to see  that if we
subtract all integrals in the perturbative expansion 
of $T(p)$ and after sum  these subtracted series we will get the
non-perturbative expression for the amplitude (the inverse of (\ref{2})) with
subtracted integrals.  If one takes
into account the contributions of all relevant counter-terms, then one will not
encounter any problems like the impossibility of removing the regularization for
positive effective range.
    
As far as  renormalised amplitude contains
contributions
of an {\it infinite} number of counter-terms with an increasing (up to {\it
infinity}) number of derivatives and consequently {\it it does not
correspond to any
 zero-range potential} the condition (\ref{cut12}) can not keep
the effective range non-positive.

\medskip
\medskip

Below we will proceed by performing subtractions without specifying
regularization. 
Subtracting divergent integrals at
$p^2=-\mu^2$ in (\ref{2}) we get the following expression:
$$
{1\over T(p)}={1\over C^R(\mu )+2p^2C_2^R(\mu )}+{M\over
4\pi}\mu+{M\over 4\pi}ip
$$
\begin{equation}
=
-{M\over 4\pi}\left\{ {1\over -{M\over 4\pi}
\left(C^R(\mu )+2p^2C_2^R(\mu )\right)}-\mu -ip\right\}
\label{4}
\end{equation}
where $C^R(\mu )$ and $C_2^R(\mu )$ are renormalised coupling constants.

We fix $C^R(\mu )$ and $C_2^R(\mu )$ by fitting (\ref{4}) to the effective range
expansion (\ref{cut5}).
For $C^R$ and $C_2^R$ we have:
\begin{equation}
C^R={4\pi a\over M(1-a\mu )}; \ \ \ \ C_2^R={\pi\over M(1-a\mu )^2}a^2r_e
\label{6}
\end{equation}
Note that above we did not use any regularization at all, so if one uses 
the dimensional or cut-off (or any other) regularization and subtracts all
integrals (takes into account the
contributions of all counter-terms which are required by QFT approach) one gets
the same results.

The expression (\ref{4}) can describe positive as well as 
negative effective  range.

The expansion parameter in (\ref{4}) (if expanded in powers of $p^2$) is: 
\begin{equation}
\lambda ={r_ep^2\over 2\left( 1/a-\mu\right)}
\label{exppar}
\end{equation}
If we take $\mu =0$ then 
\begin{equation}
\lambda ={ar_ep^2\over 2}
\label{exppar1}
\end{equation}
As considered in \cite{kaplan}, the result (\ref{exppar1}) is very discouraging
from
the EFT point of view. From (\ref{exppar1}) we see that the 
expansion of $pcot\delta ( {p})=ip+{4\pi\over M}{1\over T}$ has the radius of
convergence ${p}^2\sim 1/(ar_0)$.
But $a\sim -1/(8 MeV)$ for the 
${ }^1S_0$ channel and in general $a$ blows up as a bound state (or nearly bound
state) approaches threshold.  

It is clear from (\ref{exppar}) that the above problem has a quite natural and
simple solution. One just needs to exploit the freedom of choice of
normalization point $\mu$.
For large $a$  (\ref{exppar}) becomes $\lambda\approx {r_ep^2\over -2\mu}$.
If we take
$\mu\sim p$ we get $\lambda\sim -r_ep/2$ and the radius of convergence for 
${ }^1S_0$  is $\sim 2/r_e\approx 146 MeV$.

If we take  $\mu=140 MeV$ in (\ref{6}) we get $C^R=-{1/(99 MeV)^2}$
and $\Lambda^2 \approx (147 MeV)^2$.  So, we see that $\Lambda\sim m_{\pi}$
as was
expected for the effective theory, where pions were integrated out. So,
the effective range expansion works quite well in this problem. 
For $a\to\infty$ we have $\Lambda^2=-{r_e\over 2\mu}\sim -(140 MeV)^2$ if
$\mu\sim 140 MeV$.
Note  that
although this value of $\mu $ is not of the order of external momenta
it does not lead to any power counting problems here.
 
While this paper was in preparation (the abstract was submitted to the workshop)
the same solution of the problem of unnaturally large
scattering length's was suggested
in \cite{kaplan2}. The authors of \cite{kaplan2} used dimensional
regularization. As far as dimensional regularization discards power-law
divergences the expression for scattering amplitude is already finite in
dimensional regularization. To exploit the freedom of the choice of
normalization point one needs to perform finite subtractions. The authors of
\cite{kaplan2} called it an 'unusual subtraction scheme'. As was shown above
this solution of unnaturally large scattering length is regularization
independent
and normalization condition used is very typical and quite natural.

{\bf Conclusions}

One should be careful performing calculations in the EFT approach.
If one follows the way described in \cite{weinberg1}, \cite{weinberg2} then one
will not encounter the fundamental problems described in
\cite{cohen1,phillips1,scaldeferri,unp,beane}.
One can introduce regularization into the potential in the effective theory of
nucleons alone (this potential is taken up to
some order in EFT expansion) and solve
Schr\" oedinger equation using this regulated potential. One can fit parameters
of regularized theory using the technique of cut-off effective
theory. Otherwise, if the regularization
is supposed to be removed then it is {\it necessary} to
include contributions of an infinite number of counter-terms, or otherwise
subtract all divergences. The amplitude obtained without inclusion of
counter-terms evidently satisfies the constraint that the effective range 
has an upper bound which goes to zero in the removed cut0off limit. 
One can come to the incorrect
conclusion that the theory can not describe positive effective range. 
The analogous problems appear in the theory with included pionic (and other)
degrees of freedom and they should be treated in a similar way.

The problem of unnaturally large scattering length of
${ }^1S_0$ wave can be solved by appropriate choice of normalisation point
within conventionally renormalised theory. This solution  does not depend on
regularization.     
\medskip
\medskip
\medskip
\medskip

{\bf ACKNOWLEDGEMENTS}

I would like to thank B.Blankleider and A.Kvinikhidze for useful discussions.

This work was carried out whilst the author was a recipient of an 
 Overseas Postgraduate
Research Scholarship and a Flinders University Research Scholarship
holder at Flinders University of South Australia.

%\end{references}
\end{document}